\documentclass[conference]{IEEEtran}

\usepackage{cite}
\usepackage{amsmath,amssymb,amsfonts}
\usepackage{algorithmic}
\usepackage{graphicx}
\usepackage{textcomp}
\usepackage{xcolor}
\usepackage{multirow}  
\def\BibTeX{{\rm B\kern-.05em{\sc i\kern-.025em b}\kern-.08em
    T\kern-.1667em\lower.7ex\hbox{E}\kern-.125emX}}
\begin{document}

\title{A Lightweight Slot-Attention Framework for Multi-Instrument Multi-Pitch Estimation}

\author{\IEEEauthorblockN{Michael Taenzer}
\IEEEauthorblockA{\textit{Electronic Media Technology Group} \\
\textit{Ilmenau University of Technology}\\
Ilmenau, Germany \\
michael.taenzer@tu-ilmenau.de}
}

\maketitle

\begin{abstract}
Multi-pitch estimation (MPE) typically predicts which pitches are active in a mixture, but not which instrument or source produced them. This paper investigates a lightweight slot-attention framework for multi-instrument MPE (MI-MPE), where a mixture CQT is mapped to an unordered set of source-like pitch maps. The model uses permutation-invariant Hungarian matching to avoid fixed output semantics and treats the number of slots as an upper bound on the number of active sources. We further study two modular extensions: a self-supervised timbre encoder that provides training-time targets for slot-level timbre embeddings, and a polyphony branch that regularizes the pitch density of mixture- and slot-level predictions. Experiments show that Hungarian matching substantially improves instrument family decomposition on URMP. Stem-level prediction remains more challenging: timbre and polyphony supervision improve selected configurations, but do not consistently resolve source assignment. The results suggest that slot-based architectures are a promising direction for source-aware MPE, while highlighting the need to couple auxiliary musical cues to slot identity more carefully.
\end{abstract}

\begin{IEEEkeywords}
multi-pitch estimation, multi-instrument, slot attention, polyphony, source decomposition.
\end{IEEEkeywords}

\section{Introduction}
Music Information Retrieval (MIR) addresses tasks concerned with extracting structured information from music audio, including instrument recognition, beat tracking, source separation, transcription, and similarity estimation. Multi-pitch estimation (MPE) focuses on detecting active fundamental frequencies, typically at the frame level, and is an important step toward automatic music transcription (AMT)~\cite{mpe:christensen:2008}. In most MPE settings, predictions are made at the mixture level~\cite{mpe:bittner:2022, mpe:weiss:2022, mpetimbretrap:cwitkowitz:2024}: the model estimates which pitches are active but not which instrument or source is responsible for each pitch.

Despite substantial progress in deep learning-based methods, MPE remains difficult in fast passages, dense polyphonic textures, and signals with overlapping harmonic content. It is further complicated by ambiguous note boundaries caused by long attacks, releases, reverb, delay, or other production effects. These challenges become more pronounced in multi-instrument multi-pitch estimation (MI-MPE), where a system must not only detect active pitches, but also assign them to individual sound sources. This requires source discovery, stream assignment, and transcription at the same time. A further practical limitation is the scarcity of large open datasets~\cite{wilkins:timbredisentangle:2024} that provide isolated stems together with reliable pitch annotations, especially outside traditional acoustic instrument settings.

In this paper, we address MI-MPE with a slot-based model that predicts a set of source-like pitch maps from a mixture. Instead of assuming a fixed output order, the model produces multiple unordered slots, each intended to explain part of the mixture. This is particularly useful for electronic music, where sound sources often do not correspond to stable traditional instrument classes and may instead be described by ambiguous or role-dependent labels such as lead, pad, bass, or accompaniment. Since such roles may change within a piece, fixed output semantics are undesirable. We therefore train the model with permutation-invariant supervision based on Hungarian matching, allowing any predicted slot to be assigned to any target source during training.

To encourage source-specific slot representations beyond pitch activity alone, we introduce a self-supervised timbre encoder trained on isolated stems. The encoder produces source-level timbre embeddings, which are used as training-time targets for a timbre prediction head in the slot model. In the FiLM-conditioned variant~\cite{perez:film:2018}, the predicted timbre embedding also modulates the slot pitch decoder. The external timbre encoder is not required at inference time; it serves as a teacher during training while the final model remains lightweight.

Finally, we investigate whether explicit polyphony information can further regularize slot-based MI-MPE. A polyphony branch predicts the number of simultaneously active pitches per frame, and we evaluate both mixture-level and slot-level polyphony supervision. This encourages the model not only to detect active pitch locations but also to produce source estimates with plausible pitch density.

The contributions of this paper are twofold:
\begin{itemize}
\item We propose a lightweight slot-based architecture for MI-MPE that predicts an unordered set of source-level pitch maps from a mixture CQT and is trained with permutation-invariant supervision.
\item We investigate two modular extensions of the slot model: a self-supervised timbre encoder used as a training-time teacher for slot-level timbre embeddings, and a polyphony branch used to regularize mixture- and slot-level pitch predictions.
\end{itemize}

We frame this work as an exploratory concept study, motivated by our focus on computationally lightweight models and open datasets with isolated stems and reliable pitch annotations. Rather than relying on large pretrained audio encoders, we study whether compact slot-based architectures can provide a foundation for future real-time-capable MI-MPE systems.

\section{Related Work}
We consider related work that connects to MI-MPE through lightweight MPE, timbre-aware transcription, source assignment, and polyphony supervision.

Lightweight MPE models provide an important starting point for efficient transcription systems. Bittner et al.~\cite{mpe:bittner:2022}, for example, propose a sequence-to-sequence multitask model that jointly estimates notes, onsets, and pitch activations. Timbre-Trap~\cite{mpetimbretrap:cwitkowitz:2024} also targets low-resource MPE with a comparatively lightweight model, but uses timbre filtering to obtain instrument-agnostic pitch salience. In contrast, our work uses timbre embeddings explicitly as a training-time signal for source decomposition.

Several recent works further highlight the importance of timbre representation learning. Wilkins et al.~\cite{wilkins:timbredisentangle:2024} use multi-view self-supervision to disentangle timbre from frequency in latent audio representations. Sato et al.~\cite{sato:timbre:2026} propose timbre-based pretraining for multi-instrument AMT, learning a VAE timbre space from harmonic features and using clustered pseudo-labels to pretrain a transcription model before fine-tuning on instrument-labeled datasets. While these works motivate timbre-aware learning, our approach differs in using timbre embeddings as supervision for unordered slot-wise source representations rather than as a pretraining target for a conventional instrument-labeled AMT model.

Other methods avoid fixed instrument labels by clustering intermediate pitch representations. Tanaka et al.~\cite{mpeclustering:tanaka:2020} cluster an instrument-independent pitch representation into source-specific parts, but the number of clusters must be specified in advance. Our model also uses a fixed slot budget, but treats it as a capacity (an upper bound): activity and inactivity supervision, combined with Hungarian matching, allow unused slots to be suppressed when fewer sources are present.

Slot attention has previously been applied to audio in AudioSlots~\cite{reddy:audioslots:2023}, where a mixture spectrogram is represented as an unordered set of source-like latent objects for blind source separation. We adopt the same object-centric view of audio mixtures, but use slots to predict source-level pitch activation maps rather than reconstruct separated spectrograms.

Finally, polyphony can be obtained implicitly from MPE systems by counting predicted active pitches, or modeled explicitly as local polyphony estimation. Previous work has used explicit polyphony estimates as side information for instrument recognition, MPE post-processing, or multitask MPE training~\cite{polyphony:kareer:2018, lpe:taenzer:2021, mpeppi:taenzer:2025, mpe:weiss:2022}. We adopt this explicit view and use frame-wise mixture and slot polyphony as auxiliary supervision to regularize the pitch density of slot-wise predictions.

\section{Methodology}
\begin{figure*}[]
\centering
\includegraphics[width=0.99\textwidth]{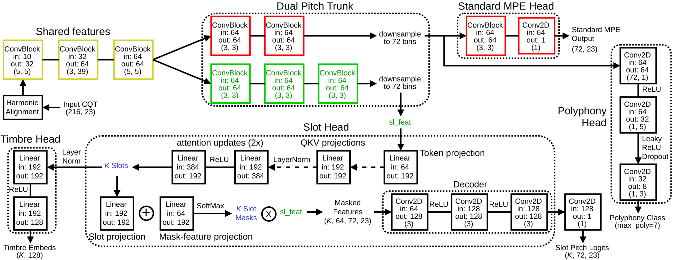}
\caption{Main slot model architecture with shared input features, standard MPE head, slot head, timbre head, and polyphony head.}
\label{fig:slot-model}
\end{figure*}

Our goal is to develop a lightweight model for MI-MPE that predicts source-level pitch maps from a mixture without assigning fixed output channels to predefined instruments. We formulate the output as an unordered set of slot-wise pitch maps. The number of slots is fixed but interpreted as an upper bound on the number of active sources; unused slots are suppressed through activity and inactivity supervision. The model consists of a slot-based MPE network and two optional extensions. A separately trained timbre encoder provides source-level embeddings for supervising a slot-level timbre head; in the FiLM variant, the predicted slot embedding also conditions the slot pitch decoder. A polyphony branch provides auxiliary frame-wise supervision for the number of active pitches at the mixture and slot level. The experimental configurations in Table~\ref{tab:experiments-overview-small} and Section~\ref{sec:experiments} add these components step by step. An architecture overview is presented in Fig.~\ref{fig:slot-model}.

\subsection{Input Representation}
All models operate on constant-Q transforms (CQTs) computed from 16 kHz mono audio. We use 36 bins per octave over six octaves, resulting in 216 frequency bins, or 72 semitone pitches with three bins per semitone. The hop size is 512 samples, corresponding to a frame resolution of approx. 32 ms. Pitch targets are restricted to the 72-bin range from MIDI 23 to 95. To obtain an efficient HCQT-like representation, we use the fast-HCQT approximation from~\cite{balhar:fasthcqt:2018}, also used in~\cite{mpe:bittner:2022}: For each selected harmonic, the CQT is shifted so that harmonic energy is aligned with candidate fundamentals. The shifted harmonic channels are concatenated with corresponding validity masks. With five harmonics and one subharmonic, this yields a 12-channel aligned representation. Ultimately, the input to the convolutional frontend has a shape of $\mathbb{R}^{12 \times 216 \times 23}$.

\subsection{Slot-Based MI-MPE Model}

The aligned representation is passed through a convolutional trunk inspired by Basic Pitch~\footnote{https://github.com/spotify/basic-pitch}. The trunk first operates at the original 216-bin resolution and then splits into two branches: The global branch predicts a mixture pitch activation map, while the slot branch produces features for family- or stem-level slot prediction. Both branches are reduced from 216 to 72 bins by averaging over groups of three bins.

The slot branch is processed by a slot head: Its feature map is flattened over frequency and time into pitch-time tokens, which are projected into the slot dimension ($d=192$). A fixed set of learned slot queries then iteratively attends to these tokens. Attention is normalized over slots for each token,

\begin{equation}
    a_{kj} =
    \frac{\exp(q_k^\top z_j / \sqrt{d})}
    {\sum_{r=1}^{K} \exp(q_r^\top z_j / \sqrt{d})},
\end{equation}

where $a_{kj}$ is the attention weight assigning pitch-time token $j$ to slot $k$. The vector $q_k$ is the query produced from slot $k$, $z_j$ is the key produced from input token $j$. Since the denominator sums over all $K$ slots, the slots compete for ownership of each pitch-time token. After each attention step, the slot states are updated by aggregating token values and applying residual feed-forward refinement. The final $K$ slot vectors represent an unordered set of latent source hypotheses.

Each slot is decoded into a pitch activation map with a mask-based decoder. Slot vectors and pitch-time features are projected into a shared space, and their similarity defines soft masks over the feature map. These masks are normalized across slots, encouraging different slots to explain different regions of the representation. A slot activity probability predicts whether each slot is active, and gates each mask before decoding. The masked feature map is then passed through a small convolutional decoder to obtain one pitch map per slot, yielding logits in $\mathbb{R}^{K \times 72 \times 23}$. These are also combined into a global pitch estimate via sigmoid and a noisy-OR operation,

\begin{equation}
    M_{\mathrm{global}} = 1 - \prod_k (1 - M_k)
\end{equation}

where $M_k$ denotes the pitch probability map predicted by slot $k$. This provides a consistency signal between the unordered set of slot predictions and the mixture pitch target.

\subsection{Timbre Encoder and Timbre Conditioning}
The timbre encoder is trained separately on isolated stem CQT windows and is used as a teacher for training the slot model. It applies the same harmonic alignment, followed by a convolutional backbone, adaptive mean/max pooling, and a fully connected network that outputs a 128-dimensional stem-timbre embedding. A separate projection head maps this embedding to a lower-dimensional space used only during contrastive training. The full encoder has 699,584 trainable parameters.

It is trained with a multi-view source-identity objective. For each stem, four active windows are sampled and treated as positive views of the same source; windows from other sources in the batch act as negatives. Window selection is activity-aware using pitch labels or CQT energy as a fallback. The main loss is a supervised multi-view contrastive loss similar to NT-Xent~\cite{chen:ntxent:2020}, combined with within-source consistency, variance, and covariance regularization. After training, the encoder is frozen. Several active windows are embedded per stem, averaged, normalized, and stored as one source-level timbre target. During slot-model training, each predicted slot embedding is supervised against the corresponding source embedding after Hungarian matching. In the FiLM-conditioned variant, the predicted slot timbre embedding modulates the masked slot feature map before pitch decoding. This external timbre encoder is not required at inference time.

\subsection{Training Objective}

The model is trained with a weighted sum of losses, with different experiments activating different subsets of terms. All pitch-map losses use binary cross-entropy (BCE). Experiment (Exp) 1a uses only a mixture MPE loss between global pitch logits and the mixture pitch target. Exp~1b adds a fixed-order slot pitch loss against pre-defined family targets as well as a slot-union loss, which compares the noisy-OR combination of all slot predictions to the global pitch target, encouraging the set of slots to jointly explain the mixture pitch activity.

For permutation-invariant experiments, predicted slots are matched to target family/stem pitch maps using Hungarian matching. The matching cost is based on BCE between predicted and target pitch maps; in timbre-aware experiments, cosine distance between predicted and target timbre embeddings is added to this cost. After matching, the slot pitch loss is computed only between assigned slot-target pairs. The same assignment also determines which predicted timbre embedding is compared to which precomputed source embedding. Matched slots are treated as active and unmatched slots as inactive. We therefore add a BCE activity loss on the scalar slot activity logits and an inactive-slot pitch loss that penalizes pitch activations in unmatched slots.

In Exp~2 and Exp~3, matched slot timbre embeddings are supervised with a cosine-distance loss against the precomputed source embeddings. In Exp~3, global polyphony is supervised with frame-wise cross-entropy over clipped polyphony classes. Counts are capped at seven, so frames with more than seven active notes are mapped to the highest class. For slot-level polyphony supervision, the pitch probability mass of each matched slot is summed over frequency and compared to the corresponding source polyphony target using a Smooth L1 loss. The contribution of each term is controlled by empirically chosen weights, kept consistent within experiment groups.

\section{Experiments}
\label{sec:experiments}

\begin{table}[]
\caption{Overview of the experimental configurations. Hung. denotes Hungarian matching, Mix and Slot denote mixture- and slot-level polyphony losses.}
\label{tab:experiments-overview-small}
\resizebox{\columnwidth}{!}{%
\begin{tabular}{c|c|c|cc|ccc|c}
\multirow{2}{*}{Exp} & \multirow{2}{*}{Task} & \multirow{2}{*}{Slots} & \multicolumn{2}{c|}{Timbre} & \multicolumn{3}{c|}{Polyphony} & \multirow{2}{*}{\begin{tabular}[c]{@{}c@{}}Model\\ Params\end{tabular}} \\ 
 &  &  & \multicolumn{1}{c|}{Head} & FiLM & \multicolumn{1}{c|}{Head} & \multicolumn{1}{c|}{Mix} & Slot  &  \\ \hline
1a & standard & - & \multicolumn{1}{c|}{-} & - & \multicolumn{1}{c|}{-} & \multicolumn{1}{c|}{-} & - & 1,266,211 \\
1b & family & 4, fixed & \multicolumn{1}{c|}{-} & - & \multicolumn{1}{c|}{-} & \multicolumn{1}{c|}{-} & - & 1,266,211 \\
1c & family & 4, Hung. & \multicolumn{1}{c|}{-} & - & \multicolumn{1}{c|}{-} & \multicolumn{1}{c|}{-} & - & 1,266,211 \\ \hline
2a & stem & 8, Hung. & \multicolumn{1}{c|}{\checkmark} & - & \multicolumn{1}{c|}{-} & \multicolumn{1}{c|}{-} & - & 1,329,123 \\
2b & stem & 8, Hung. & \multicolumn{1}{c|}{\checkmark} & \checkmark & \multicolumn{1}{c|}{-} & \multicolumn{1}{c|}{-} & - & 1,378,851 \\ \hline
3a & stem & 8, Hung. & \multicolumn{1}{c|}{\checkmark} & \checkmark & \multicolumn{1}{c|}{\checkmark} & \multicolumn{1}{c|}{-} & - & 1,684,875 \\
3b & stem & 8, Hung. & \multicolumn{1}{c|}{\checkmark} & \checkmark & \multicolumn{1}{c|}{\checkmark} & \multicolumn{1}{c|}{\checkmark} & - & 1,684,875 \\
3c & stem & 8, Hung. & \multicolumn{1}{c|}{\checkmark} & \checkmark & \multicolumn{1}{c|}{\checkmark} & \multicolumn{1}{c|}{\checkmark} & \checkmark & 1,684,875
\end{tabular}%
}
\end{table}

We evaluate the model through a controlled series of experiments that incrementally extend standard MPE toward slot-based stem-level prediction with timbre and polyphony supervision. Since our goal is to study the feasibility and behavior of slot-based MI-MPE under lightweight constraints, the evaluation focuses on controlled internal ablations rather than comparison to large-scale transcription systems. Table~\ref{tab:experiments-overview-small} summarizes the configurations.

Exp~1 introduces the slot formulation, and starts with a standard MPE baseline (1a). Exp~1b adds four fixed family-level slots for the traditional instrument families brass, keyboards, strings, and woodwinds. Exp~1c replaces the fixed slot order with Hungarian matching between predicted slots and active family targets, and adds activity and inactivity supervision.

Exp~2 moves from family-level to stem-level prediction by increasing the number of slots to eight. Exp~2a adds a timbre prediction head; Exp~2b additionally uses the predicted timbre embedding for FiLM conditioning of the slot pitch decoder.

Exp~3 keeps the same stem-level setup and adds polyphony-related components. Exp~3a adds the polyphony head while disabling its losses, isolating the architectural effect of the branch. Exp~3b adds global polyphony supervision derived from the mixture-level pitch target, and Exp~3c further adds slot-level polyphony supervision after Hungarian matching.

\subsection{Datasets}
We focus primarily on the URMP~\cite{URMP:2019} and mshoxxDB~\cite{mpeppi:taenzer:2025} datasets, because they provide both mixture audio and isolated stems with respective $f_0$ / pitch annotations. The two datasets are complementary: URMP contains acoustic ensemble recordings with traditional instruments, while mshoxxDB contains electronic music with diverse synthetic and processed timbres. We additionally report MusicNet~\cite{musicnet:2017} only for the standard MPE baseline in Exp~1a, to relate the performance of our model to prior work on a widely used benchmark.

\subsubsection{URMP}
The University of Rochester Multi-modal Music Performance (URMP) Dataset~\cite{URMP:2019} contains 44 ensemble arrangements of duets, trios, quartets, and quintets, mostly based on popular classical repertoire. It provides multitrack recordings with 149 isolated instrument tracks. We use the stems both directly and grouped by instruments into traditional families based on the instrument codes in the filenames, as shown in Table~\ref{tab:instrument-families-urmp}. Since URMP has no official split, we use a custom 30/7/7 train/val/test split considering the number of instruments per piece so that val and test each contain at least two pieces with three or more instruments.

\begin{table}[]
\caption{URMP instrument code to instrument family assignment.}
\label{tab:instrument-families-urmp}
\centering
\begin{tabular}{c|c|c}
Code & Instrument & Instrument Family \\ \hline
Vn. & Violin & Strings \\
Va. & Viola & Strings \\
Vc. & Cello & Strings \\
Db. & Double Bass & Strings \\
Fl. & Flute & Woodwinds \\
Ob. & Oboe & Woodwinds \\
Cl. & Clarinet & Woodwinds \\
Sax. & Saxophone & Woodwinds \\
Bn. & Bassoon & Woodwinds \\
Tpt. & Trumpet & Brass \\
Hn. & Horn & Brass \\
Tbn. & Trombone & Brass \\
Tba. & Tuba & Brass
\end{tabular}
\end{table}

\subsubsection{mshoxxDB}
This dataset focuses on electronic music and covers subgenres such as 8-bit (chiptune), house, and dreamy styles. It consists of 18 pieces with 125 isolated tracks (including drum stems), and approx. 61 minutes of audio. Because it combines diverse synthetic timbres, processed sounds with additional effects (e.g. delays and reverbs) and some classical instruments, it provides a challenging setting for MI-MPE and is particularly useful for training and evaluating timbre-aware components. We use the \texttt{ms12} split proposed in~\cite{mpeppi:taenzer:2025}.

\subsubsection{MusicNet}
MusicNet is a collection of classical music recordings of solo performances and small instrumental ensembles. It comprises 330 pieces and 34.1 hours of music with annotations for 11 instruments. We use these annotations to derive frame-level mixture pitch targets for the standard MPE baseline in Exp~1a. Although the labels could support further slot experiments, MusicNet does not provide isolated source audio and is therefore not used for the stem protocol. We use the \texttt{MuN-10} split proposed in~\cite{mpe:weiss:2022}.

\section{Training \& Evaluation Protocol}
\subsection{Training Parameters}
All experiments use windows of 23 CQT frames, corresponding to 736 ms, and predict output sequences of the same length. During training, validation loss is computed on deterministic fixed-length validation windows of the same duration. The slot model is trained with the AdamW optimizer using an initial learning rate of $10^{-3}$, reduced by factor 0.5 when total validation loss plateaus, and a batch size of 8. Checkpoints are selected primarily by total validation loss; relevant auxiliary losses are considered for timbre- and polyphony-aware experiments.

The timbre encoder is also trained with AdamW, a starting learning rate of $5 \times 10^{-4}$, cosine annealing, and a batch size of 32. Each self-supervised training item consists of four active windows from the same isolated source, each 46 frames long (approx. 1.5 s). Activity-aware sampling uses pitch labels when available and CQT energy as a fallback, which is important for mshoxxDB where long releases, reverb, or delay can make pitch labels unreliable. Mild augmentation is applied through random gain between -3 and 3 dB, random frequency/time masking, and window normalization.

Both models are trained for a maximum of 100 epochs with an early stopping patience of 6.

\subsection{Evaluation}
Evaluation is performed on full files rather than fixed-length windows: Prediction thresholds are selected using full-file validation inference with batch size 1 and then fixed for full-file test inference. We report frame-level MPE metrics: average precision (AP), precision (P), recall (R), and F1-score. Metrics are computed for mixture-level predictions and, where applicable, for matched family or stem slot predictions. Slot outputs are matched to target maps with Hungarian matching before stem evaluation. For the auxiliary components, we report the mean cosine similarity (cos mean) between matched predicted and target timbre embeddings, as well as global and source-level polyphony accuracy (standard acc, stem acc).

All experiments were run on a Ryzen 4500U system with 16 GB RAM and no dedicated GPU. On the largest configuration (Exp 3), training one epoch on URMP takes approx. 9 seconds. This setup reflects the intended focus on lightweight models that remain feasible under limited computational resources.

\section{Results and Discussion}
\label{sec:results}

\begin{table*}[]
\caption{Main evaluation results on the individual test sets. Family MPE in Exp 1, Stem MPE in Exp 2 \& 3. Parentheses denote polyphony head is present but not supervised. All metrics reported as percentages, except cosine similarity.}
\label{tab:results}
\resizebox{\textwidth}{!}{%
\begin{tabular}{c|c|cccc|cccc|c|cc}
\multirow{2}{*}{Exp} & \multicolumn{1}{c|}{\multirow{2}{*}{Dataset}} & \multicolumn{4}{c|}{Standard MPE} & \multicolumn{4}{c|}{Family / Stem MPE} & Timbre & \multicolumn{2}{c}{Polyphony} \\
 & \multicolumn{1}{c|}{} & AP $\uparrow$ & F1 $\uparrow$ & P $\uparrow$ & R $\uparrow$ & AP $\uparrow$ & F1 $\uparrow$ & P $\uparrow$ & R $\uparrow$ & cos mean $\uparrow$ & standard acc $\uparrow$ & stem acc $\uparrow$ \\ \hline
1a & MusicNet & 60.31 & 58.03 & 48.90 & 71.35 & - & - & - & - & - & - & - \\
1a & mshoxxDB & 21.93 & 29.27 & 23.10 & 39.95 & - & - & - & - & - & - & - \\
1a & URMP & 74.78 & 69.08 & 73.20 & 65.39 & - & - & - & - & - & - & - \\
1b & URMP & 76.40 & 70.67 & 71.89 & 69.50 & 24.00 & 33.91 & 26.97 & 45.67 & - & - & - \\
1c & URMP & 77.27 & 70.50 & 70.98 & 70.03 & 61.12 & 65.91 & 59.70 & 73.56 & - & - & - \\ \hline
2a & mshoxxDB & 41.06 & 41.65 & 34.32 & 52.97 & 9.31 & 10.58 & 6.15 & 37.80 & 0.7404 & - & - \\
2b & mshoxxDB & 39.93 & 42.23 & 36.72 & 49.68 & 19.81 & 18.57 & 12.06 & 40.33 & 0.6940 & - & - \\
2a & URMP & 73.31 & 68.25 & 68.87 & 67.65 & 24.94 & 33.78 & 24.57 & 54.07 & 0.8452 & - & - \\
2b & URMP & 75.51 & 70.72 & 72.21 & 69.30 & 30.76 & 39.08 & 29.62 & 57.43 & 0.8902 & - & - \\ \hline

3a & mshoxxDB & 43.27 & 43.97 & 38.61 & 51.07 & 13.41 & 18.87 & 11.79 & 47.22 & 0.6514 & (8.19) & (44.04) \\
3b & mshoxxDB & 41.12 & 44.51 & 41.37 & 48.16 & 7.72 & 13.79 & 8.26 & 41.81 & 0.5814 & 30.12 & 8.89 \\
3c & mshoxxDB & 43.72 & 44.44 & 42.62 & 46.42 & 13.08 & 15.62 & 9.20 & 51.66 & 0.7430 & 30.21 & 25.29 \\
3a & URMP & 71.85 & 68.29 & 71.14 & 65.66 & 8.83 & 13.15 & 8.00 & 36.87 & 0.5629 & 13.36 & 35.45 \\
3b & URMP & 72.01 & 66.89 & 68.97 & 64.94 & 33.16 & 42.17 & 33.04 & 58.27 & 0.8577 & 29.34 & 69.39 \\
3c & URMP & 74.20 & 69.08 & 69.61 & 68.57 & 35.43 & 42.60 & 40.90 & 44.45 & 0.8537 & 31.95 & 60.08
\end{tabular}%
}
\end{table*}

Table~\ref{tab:results} summarizes the main results. Exp~1a provides the standard MPE baseline. On MusicNet, it reaches an AP of 60.31, below the approx. 75 AP reported in~\cite{mpe:weiss:2022, lpe:taenzer:2021}. Given the deliberately compact architecture and the focus on controlled slot-model extensions, this provides a reasonable starting point for the remaining experiments. mshoxxDB and URMP achieve a baseline AP of 21.93 and 74.78 and F1 of 29.27 and 69.08, respectively. This already highlights that mshoxxDB is a substantially harder MPE setting (see Section~\ref{sec:experiments}). On URMP, the permutation-invariant formulation in Exp~1c substantially improves over the fixed slots in Exp~1b, with family AP increasing from 24.00 to 61.12 and family F1 from 33.91 to 65.91. This confirms that fixed semantic slot assignments are too restrictive, and that Hungarian matching is important for meaningful slot specialization. At the same time, standard MPE also improves, suggesting that the auxiliary slot task can benefit the shared representation rather than merely adding source-level outputs.

\begin{figure}[]
\centering
\includegraphics[width=0.99\linewidth]{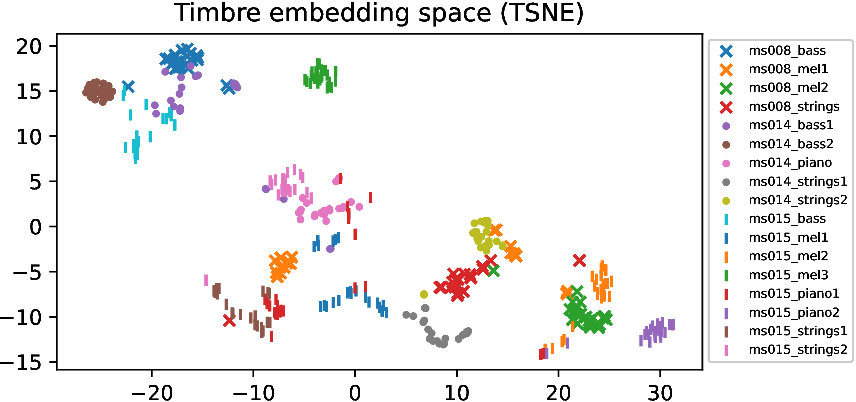}
\caption{The timbre encoder produces grouped stem clusters (\texttt{ms12}-test). Occasional overlap with other clusters due to timbre similarity. The legend allows for a rough comparison of file groups and instrument classes.}
\label{fig:timbre-space}
\end{figure}

The subsequent experiments require timbre embeddings. The checkpoint of the trained timbre encoder was selected using fixed-set top-1 same-source retrieval accuracy on active validation windows. It achieved 71.52\% top-1 and 92.36\% top-5 retrieval accuracy, indicating that the embeddings consistently group windows from the same source while remaining non-collapsed; see Fig.~\ref{fig:timbre-space} for examples.

As expected, stem-level MPE in Exp~2 makes the task considerably harder. On URMP, FiLM-based timbre conditioning in Exp~2b improves over Exp~2a: standard F1 increases from 68.25 to 70.72, stem F1 from 33.78 to 39.08, and matched timbre cosine similarity from 84.52 to 89.02. On mshoxxDB, the effect is more mixed: Exp~2b improves stem AP from 9.31 to 19.81 and stem F1 from 10.58 to 18.57, but slightly decreases standard AP and matched timbre cosine similarity. Thus, timbre conditioning helps stem attribution in both datasets, but its effect is less stable in the more heterogeneous mshoxxDB setting. Interestingly, the timbre head alone improves global AP on mshoxxDB considerably, but slightly reduces it for URMP. This suggests a trade-off between stem-structuring objectives and the shared pitch representation, discussed in Section~\ref{sec:conclusions-limitations}.

Adding the polyphony head without activating its losses in Exp~3a appears harmful for URMP stem prediction, but explicit polyphony supervision recovers and even improves performance: Exp~3c reaches the best URMP stem AP and F1 among all stem-level experiments, with 35.43 and 42.60. This suggests that slot-level polyphony can help constrain source predictions. However, on mshoxxDB, the polyphony variants do not clearly improve over Exp~2b; Exp~3c improves standard AP and timbre cosine similarity, but stem-level performance remains below Exp~2b. Overall, the results show: permutation-invariant slot supervision is essential, and timbre and polyphony provide useful but dataset-dependent regularization signals.

\section{Conclusions and Limitations}
\label{sec:conclusions-limitations}
This paper presents a lightweight slot-based approach to MI-MPE. The results show that unordered slot-wise pitch prediction is viable, but only when slot ambiguity is handled explicitly through permutation-invariant matching, active-target filtering, slot activity supervision, and inactive-slot suppression. The strongest gains are obtained with permutation-invariant prediction, suggesting that fixed output semantics are too restrictive for source decomposition.

Source assignment remains challenging. Timbre and polyphony supervision provide useful musical cues, but do not automatically yield cleaner source-level pitch masks. This suggests that auxiliary structure should primarily resolve slot identity or pitch-density ambiguity, rather than act as an unconstrained prediction head or decoder conditioning signal.

A central limitation is therefore the coupling between auxiliary objectives and slot decomposition: timbre supervision can improve slot-embedding alignment without improving the corresponding pitch masks, and timbre-conditioned decoding may introduce timbre-dependent pitch priors instead of simply stabilizing slot identity. Future work should explore more disentangled uses of these cues~\cite{disentangle:laura:2026}, adaptive slot allocation, stronger duplicate suppression, time-varying timbre embeddings, and scaling to larger datasets such as Slakh2100~\cite{slakh:manilow:2019}.

\bibliographystyle{IEEEtran}
\bibliography{references}

\end{document}